\documentclass[conference]{IEEEtran}
\IEEEoverridecommandlockouts

\usepackage{cite}
\usepackage{amsmath,amssymb,amsfonts}
\usepackage{algorithmic}
\usepackage{graphicx}
\usepackage{textcomp}
\usepackage{xcolor}
\usepackage{caption} 
\usepackage{float}
\usepackage{url}
\usepackage{array}

\def\BibTeX{{\rm B\kern-.05em{\sc i\kern-.025em b}\kern-.08em
    T\kern-.1667em\lower.7ex\hbox{E}\kern-.125emX}}
\begin{document}



\title{HER2 and FISH Status Prediction in Breast Biopsy H\&E-Stained Images Using Deep Learning\\
}

\author{\IEEEauthorblockN{Ardhendu Sekhar}
\IEEEauthorblockA{\textit{Department of Electrical Engineering} \\
\textit{Indian Institute of Technology, Bombay}\\
Mumbai, India \\
asekhar@iitb.ac.in}
\and
\IEEEauthorblockN{Vrinda Goel}
\IEEEauthorblockA{\textit{Department of Electrical Engineering} \\
\textit{Indian Institute of Technology, Bombay}\\
Mumbai, India \\
20d070090@iitb.ac.in}
\and
\IEEEauthorblockN{Garima Jain}
\IEEEauthorblockA{\textit{Department of Pathology} \\
\textit{Indian Council of Medical Research}\\
Delhi, India \\
garima.j@icmr.gov.in}
\and
\IEEEauthorblockN{Abhijeet Patil}
\IEEEauthorblockA{\textit{Department of Electrical Engineering} \\
\textit{Indian Institute of Technology, Bombay}\\
Mumbai, India \\
abhijeetptl@iitb.ac.in}
\and
\IEEEauthorblockN{Ravi Kant Gupta}
\IEEEauthorblockA{\textit{Department of Electrical Engineering} \\
\textit{Indian Institute of Technology, Bombay}\\
Mumbai, India \\
184070025@iitb.ac.in}
\and
\IEEEauthorblockN{Tripti Bameta}
\IEEEauthorblockA{\textit{Computational Pathology Laboratory} \\
\textit{Tata Memorial Centre-ACTREC, HBNI}\\
Navi Mumbai, India \\
tripti.bameta@gmail.com}
\and
\IEEEauthorblockN{Swapnil Rane}
\IEEEauthorblockA{\textit{Department of Pathology} \\
\textit{Tata Memorial Centre-ACTREC, HBNI}\\
Navi Mumbai, India \\
raneswapnil82@gmail.com}
\and
\IEEEauthorblockN{ }
\IEEEauthorblockA{ } 
\and
\IEEEauthorblockN{ }
\IEEEauthorblockA{ } 
\and
\IEEEauthorblockN{ }
\IEEEauthorblockA{ } 
\and
\IEEEauthorblockN{ }
\IEEEauthorblockA{ } 
\and
\IEEEauthorblockN{ }
\IEEEauthorblockA{ } 
\and
\IEEEauthorblockN{ }
\IEEEauthorblockA{ } 
\and
\IEEEauthorblockN{ }
\IEEEauthorblockA{ } 
\and
\IEEEauthorblockN{ }
\IEEEauthorblockA{ } 
\and
\IEEEauthorblockN{ }
\IEEEauthorblockA{ } 
\and
\IEEEauthorblockN{ }
\IEEEauthorblockA{ } 
\and
\IEEEauthorblockN{ }
\IEEEauthorblockA{ } 
\and
\IEEEauthorblockN{ }
\IEEEauthorblockA{ } 
\and
\IEEEauthorblockN{ }
\IEEEauthorblockA{ } 
\and
\IEEEauthorblockN{ }
\IEEEauthorblockA{ } 
\and
\IEEEauthorblockN{ }
\IEEEauthorblockA{ } 
\and
\IEEEauthorblockN{ }
\IEEEauthorblockA{ } 
\and
\IEEEauthorblockN{ }
\IEEEauthorblockA{ } 
\and
\IEEEauthorblockN{ }
\IEEEauthorblockA{ } 
\and
\IEEEauthorblockN{ }
\IEEEauthorblockA{ } 
\and
\IEEEauthorblockN{ }
\IEEEauthorblockA{ } 
\and
\IEEEauthorblockN{ }
\IEEEauthorblockA{ } 
\and
\IEEEauthorblockN{ }
\IEEEauthorblockA{ } 
\and
\IEEEauthorblockN{ }
\IEEEauthorblockA{ } 
\and
\IEEEauthorblockN{ }
\IEEEauthorblockA{ } 
\and
\IEEEauthorblockN{ }
\IEEEauthorblockA{ } 
\and
\IEEEauthorblockN{ }
\IEEEauthorblockA{ } 
\and
\IEEEauthorblockN{ }
\IEEEauthorblockA{ } 
\and
\IEEEauthorblockN{ }
\IEEEauthorblockA{ } 
\and
\IEEEauthorblockN{ }
\IEEEauthorblockA{ } 
\and
\IEEEauthorblockN{ }
\IEEEauthorblockA{ } 
\and
\IEEEauthorblockN{Amit Sethi}
\IEEEauthorblockA{\textit{Department of Electrical Engineering} \\
\textit{Indian Institute of Technology, Bombay}\\
Mumbai, India \\
asethi@iitb.ac.in}
}

\maketitle

\begin{abstract}

The current standard for detecting human epidermal growth factor receptor 2 (HER2) status in breast cancer patients relies on HER2 expression identified through immunohistochemistry (IHC) or amplification identified through fluorescence in situ hybridization (FISH). However, hematoxylin and eosin (H\&E) tumor stains are more widely available, and accurately predicting HER2 status using H\&E could reduce costs and expedite treatment selection. Deep Learning algorithms for H\&E have shown effectiveness in predicting various cancer features and clinical outcomes, including moderate success in HER2 status prediction. In this work, we employed a customized weak supervision classification technique combined with MoCo-v2 contrastive learning for self-supervised feature extraction training to predict HER2 status. We trained our pipeline on 182 publicly available H\&E whole slide images (WSIs) from The Cancer Genome Atlas (TCGA), for which  annotations by the pathology team at Yale School of Medicine are publicly available. Our pipeline achieved an Area Under the Curve (AUC) of 0.85 ± 0.02 across four different test folds. Additionally, we tested our model on 44 H\&E slides from the TCGA-BRCA dataset, which had an HER2 score of 2+ and included corresponding HER2 status and FISH test results. These cases are considered equivocal for IHC, requiring an expensive FISH test on their IHC slides for disambiguation. Our pipeline demonstrated an AUC of 0.81 on these challenging H\&E slides. Reducing the need for FISH test can have significant implications in cancer treatment equity for underserved populations.

\end{abstract}


\section{Introduction}

Breast carcinoma holds the highest mortality rate among women globally~\cite{b1}~\cite{b2}. Treatment strategies vary depending on the well-established subtypes of breast cancer, each defined by distinct sets of genetic mutations. Among the four primary subtypes: luminal A, luminal B, HER2, and basal, HER2 was historically the most deadly until recently~\cite{b3}. The HER2 subtype is marked by the overexpression of the HER2 gene, often accompanied by mutations in other genes within the HER2 amplicon, such as GRB7~\cite{b5}, PGAP3~\cite{b4}, and, to some extent, TP53, which contribute to tumor growth~\cite{b6}. The introduction of targeted anti-HER2 therapies (e.g., trastuzumab, lapatinib, and pertuzumab) has significantly reduced mortality for HER2 positive breast cancer patients, though these treatments are costly. However, these therapies are ineffective, and sometimes even detrimental, for other breast cancer subtypes~\cite{b7}. Hence, accurately determining HER2 overexpression in breast cancers is crucial. This overexpression is typically identified through precise yet expensive fluorescence in situ hybridization (FISH) tests or, more commonly, through the more affordable but less reliable HER2neu immunostaining~\cite{b6}.

\begin{figure}[htbp]
\centering
\includegraphics[width=0.45\textwidth]{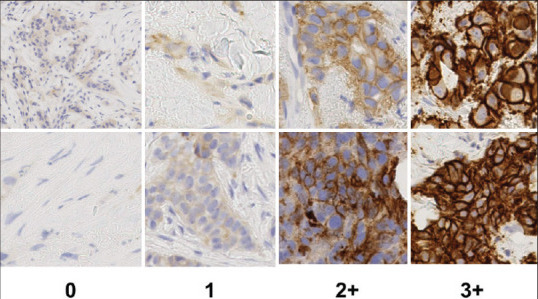}
\caption{Examples of HER2neu immunohistochemistry staining, showcasing patches from slides with different HER2 scores, each reflecting varying staining intensities.~\cite{b9}}
\label{fig:ihc_patches}
\end{figure}

HER2 is a protein that promotes cell growth and is found on the cell membrane. The evaluation of HER2 status through immunohistochemistry (IHC) is conducted in accordance with the guidelines set by the American Society of Clinical Oncology/College of American Pathologists (ASCO/CAP)~\cite{b8}. In this process, pathologists examine tissue samples that have been formalin-fixed and paraffin-embedded, with HER2 IHC staining, under a microscope. Hematoxylin counterstaining is used to make the nuclei appear faint blue. In cells that are HER2 negative, the HER2 IHC stain exhibits little to no brown coloration around the cell membrane. In contrast, in HER2 positive cells, a brown boundary encircles the nucleus, forming a chicken-wire pattern that signifies the presence of the HER2 protein. Pathologists score these stained tissue slides by visually inspecting them under a microscope, following the ASCO/CAP criteria detailed in Table~\ref{tab:ihc_table}~\cite{b8} and depicted in Figure \ref{fig:ihc_patches}~\cite{b9}. However, the interpretation of these guidelines can be subjective, especially when classifying borderline cases. For example, estimating the percentage of invasive cells with a particular staining pattern or distinguishing between “faint” and “moderate” staining can vary from one pathologist to another, as illustrated in Figure ~\ref{fig:ihc_patches}~\cite{b9}. Among all breast cancer cases, approximately 10-15\% show HER2 IHC expression at a level of 2+, which is considered equivocal or ambiguous. These cases are typically sent for further confirmation using fluorescence in situ hybridization (FISH) to definitively determine HER2 gene amplification. The latter is definitive but typically more expensive and time-consuming for underserved populations, and there are serious economic consequences of inaccurate testing by IHC ~\cite{b21}~\cite{b22}. 

\begin{table}[h]
\centering
\caption{Scoring guidelines for human epidermal growth factor receptor 2 (HER2) by the American Society of Clinical Oncology/College of American Pathologists (ASCO/CAP).~\cite{b8}}
\label{tab:ihc_table}
\begin{tabular}{|c|m{4.5cm}|c|} 
\hline
\textbf{Score} & \centering\textbf{Pattern} & \textbf{Assessment} \\ \hline
0  & \centering No observable staining, or membrane staining that is incomplete and is faint/barely perceptible in $<$ 10\% of tumor cells & Negative  \\ \hline
1+  & \centering Incomplete membrane staining that is faint/barely perceptible in $>$ 10\% of invasive tumor cells	& Negative \\ \hline
2+  & \centering Circumferential membrane staining that is incomplete and/or weak/moderate in $>$ 10\% of invasive tumor cells, or complete and circumferential intense membrane staining in $<=$ 10\% of invasive tumor cells & Equivocal \\ \hline
3+ & \centering Homogeneous, dark, circumferential (chicken wire) pattern in $>$ 10\% of invasive tumor cells & Positive \\ \hline
\end{tabular}
\end{table}

Immunohistochemistry (IHC) techniques involve visually tagging tissue regions with high concentrations of specific antigens (proteins) for microscopic analysis. IHC reagents combine a tagging agent (dye or fluorescence) with an antibody that binds to the target antigen, making it a vital tool for diagnosing tumor malignancies and identifying cancer subtypes for precision medicine. On the other hand, hematoxylin and eosin (H\&E) staining is more generic, enhancing the visual contrast between basophilic nuclei and acidophilic stromal regions to reveal the tissue's spatial structure, such as the shape of nuclei and glands. While the costs and availability of various IHC panels vary depending on the targeted proteins, H\&E staining is widely accessible and cost-effective.

We developed a customized weak supervision classification technique, combined with MoCo-v2 contrastive learning, to differentiate between HER2 positive and HER2 negative breast tumors using H\&E stained sections. The training pipeline consists of three steps: extracting patches from whole slide images, using a ResNet50 encoder pre-trained with MoCo-V2 self-supervision, and training the final attention module. The TCGA-Yale dataset was employed for both training and testing, resulting in an AUC of 0.85 ± 0.02 across four different folds. Additionally, we evaluated our model on 44 H\&E slides from the TCGA-BRCA dataset, all of which had a HER2 score of 2+ and included corresponding HER2 status and FISH test results. These cases are considered equivocal, often necessitating a costly FISH test for clarification. Our pipeline achieved an AUC of 0.81 on these challenging H\&E slides.

\section{Related Work}

Several studies have explored the use of computational methods and deep learning algorithms to enhance the detection and classification of HER2 status in breast cancer, particularly using hematoxylin and eosin (H\&E) stained slides and immunohistochemistry (IHC). One significant approach was presented by Anand et al. ~\cite{b9}, who developed a multi-stage deep learning pipeline to estimate HER2 status from H\&E stained images. Their method combined a stain separation module with three convolutional neural networks (CNNs) to classify HER2-positive and HER2-
negative samples. Another approach utilized a convolutional neural network (CNN) trained on manually annotated H\&E slides, achieving high accuracy in predicting HER2 status and treatment response, suggesting that H\&E-based algorithms can effectively support clinical evaluations~\cite{b15}. In one more study, a GrayMap+ CNN model was employed to predict HER2 expression levels and gene status directly from IHC slides, showing strong agreement with pathologist assessments and high accuracy in classification~\cite{b18}. In another approach, deep learning algorithms have been applied to HER2 2+ IHC slides to predict gene amplification status, achieving moderate accuracy but highlighting the potential for triaging cases in settings where in situ hybridization testing is not readily available~\cite{b19}. Another work developed a 3-block-DenseNet deep learning model to predict HER2 expression from ultrasound images providing a non-invasive method for HER2 prediction~\cite{b20}. 

\begin{figure*}[htbp]
\centering
\includegraphics[width=0.8\textwidth]{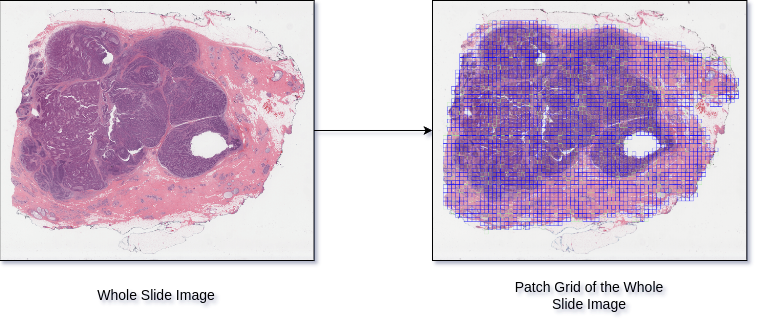}
\caption{Preprocessing of the whole slide image.}
\label{fig:patch_grid}
\end{figure*}

\begin{figure*}[htbp]
\centering
\includegraphics[width=1.05\textwidth]{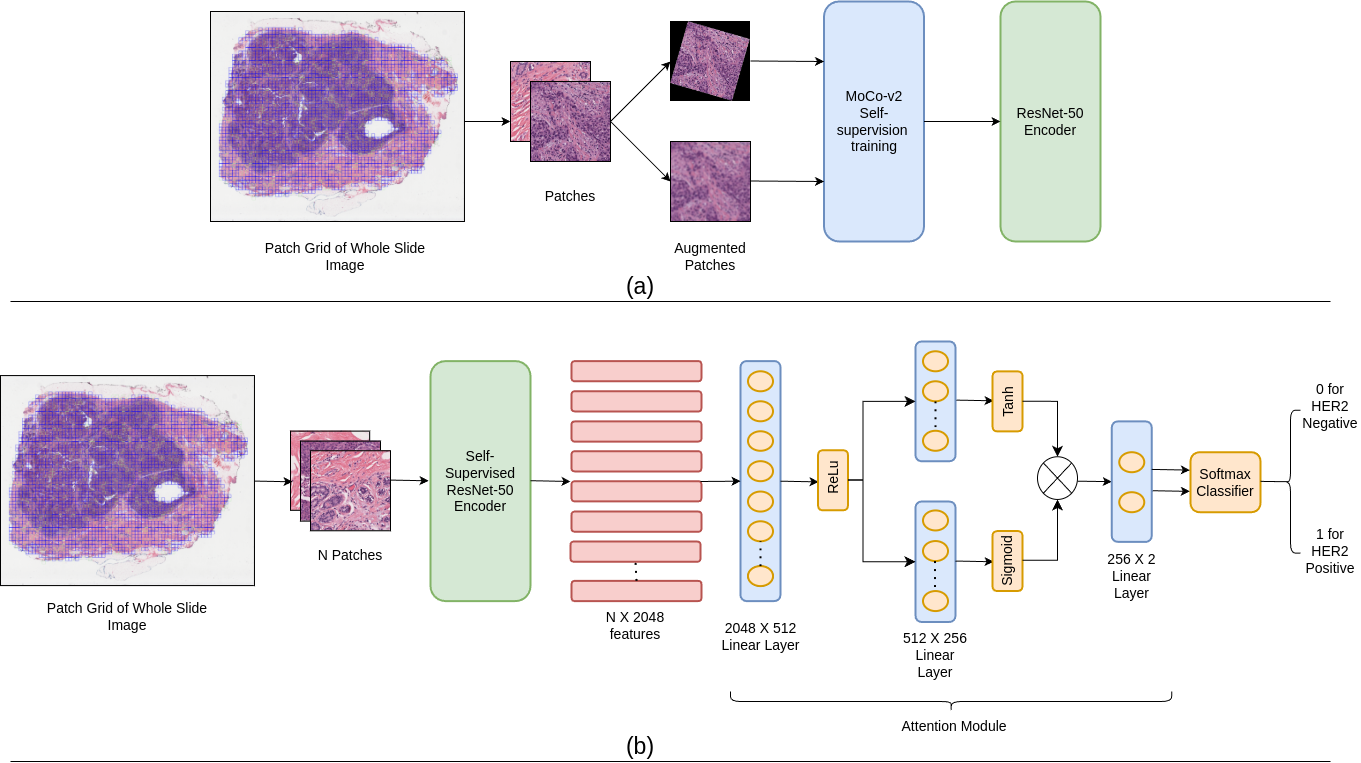}
\caption{The entire pipeline of our work is illustrated, showcasing: (a) the MoCo-v2 self-supervision training pipeline, and (b) the training and inference of the attention module.}
\label{fig:training_pipeline}
\end{figure*}

\section{Methodology}



The task at hand involves classifying whole slide images stained by inexpensive and ubiquitous hematoxylin and eosin (H\&E) stains. Because we only know the overall HER2 status of a patient but cannot localize the tissue regions showing HER2 positivity in H\&E slides, we opted for using a weakly supervised approach. Weakly supervised classification tasks in pathology typically involve training sets with known labels for each whole slide image (WSI) but lack class-specific annotations or information for individual pixels or regions. Attention-based Multiple Instance Learning (MIL)~\cite{b10} with clustering extends the MIL~\cite{b11} framework, which is well-suited for multi-class classification. Traditional MIL utilizes a non-trainable aggregation function, such as max pooling, where the slide-level prediction is determined by the patch exhibiting the highest prediction probability. In contrast, attention-based MIL with clustering employs a trainable and interpretable attention-based pooling function. This function aggregates slide-level representations from individual patch-level representations. Within the attention-based pooling mechanism, the attention network generates two distinct sets of attention scores, each corresponding to the classes in the binary classification problem. In our study, we classified breast cancer whole slide images into HER2 positive and HER2 negative categories using a weakly supervised technique proposed by Fremond et al.~\cite{b12}.

The deep learning model proposed by Fremond et al.~\cite{b12} encompasses a comprehensive approach that begins with pre-processing the whole slide images to prepare them for detailed analysis. It then trains a feature extractor model using contrastive self-supervised learning, specifically MoCo-v2~\cite{b13}, aimed at learning patch-level morphological features. Subsequently, all patch-level feature vectors are extracted from the self-supervised learning encoder. The process culminates in the training and inference of an attention-based classification model~\cite{b14}, which molecularly classifies whole slide images using the gathered patch-level feature vectors.

During the pre-processing phase, the tissue in each whole slide image was automatically segmented using Otsu thresholding. The identified tissue regions were then divided into non-overlapping square patches, each measuring $360\,\mu m$ at $40\times$ magnification, and resized to $224\times224$ pixels. To ensure quality, patches with minimal or no tissue were excluded by applying a minimum threshold of $20$ to the median value of each $8$-bit RGB channel. Additionally, patches with a mean pixel intensity across the RGB channels of approximately 245 ± 10, indicating near-white pixels, were filtered out using a specific technique\cite{b25}. A whole slide image and its corresponding patch grid, derived through the preprocessing method, are depicted in Figure~\ref{fig:patch_grid}.

To train the feature extractor model using self-supervised learning, we curated a dataset of image patches by sampling from each whole slide image. To ensure fair representation of each molecular class, we strategically sampled an optimal number of patches from whole slide images corresponding to the two classes involved in our study. This approach generated a dataset of $240,000$ tissue region image patches extracted from 120 whole slide images, evenly distributed between the two classes. We then trained MoCo-v2~\cite{b13} for 300 epochs using a ResNet-50 encoder on these whole slide image patches. MoCo-v2 is a contrastive learning framework designed to discern between similar and dissimilar data pairs, which it approaches as a dictionary lookup challenge. The learning rate was set at 0.06 with a temperature parameter of 0.07. Central to this framework is the InfoNCE loss function~\cite{b13}(equation represented below), which effectively measures the contrast between data pairs. In this setup, q denotes a query image, $x^{+}$ represents the positive (similar) key sample, and $x^{-}$ includes the negative (dissimilar) key samples. The temperature hyperparameter, $\tau$ , adjusts the sensitivity of the loss function. Within the framework's instance discrimination pretext task, a positive pair consists of a query and a key that are data-augmented versions of the same image, while a negative pair is formed when the query and key derive from different images. This methodology aids in developing features that are resilient to variations introduced by data augmentation, enhancing the model's generalization capabilities. The data augmentation techniques applied during self-supervision training included random rotations, vertical and horizontal flips, color jitter, and Gaussian Blur. 

\begin{equation}
    \mathcal{L}_{q, k^{+},\left\{k^{-}\right\}}=-\log \frac{\exp \left(q \cdot k^{+} / \tau\right)}{\exp \left(q \cdot k^{+} / \tau\right)+\sum_{k^{-}} \exp \left(q \cdot k^{-} / \tau\right)} 
\end{equation}

Feature vectors of size 2048 were obtained from each patch of the whole slide images using the self-supervised ResNet-50 encoder at its final layer. Subsequently, these vectors for all patches within a single whole slide image were aggregated. Following this, an attention-based slide classifier, derived from the CLAM~\cite{b14} architecture but without the secondary clustering objective, was trained over 100 epochs using cross entropy loss. The whole training pipeline is depicted in Figure ~\ref{fig:training_pipeline}.

\section{Experiments and Results}

In this study, we address the challenge of classifying HER2 status in breast cancer using a deep learning pipeline that leverages publicly available WSI datasets. Our approach is designed to differentiate between HER2 positive and HER2 negative tumors, as well as to classify cases with an equivocal HER2 score of 2+ into HER2 FISH positive or HER2 FISH negative categories. The following sections describe the datasets utilized in this study, the methodologies employed, and the results obtained.

\subsection{Dataset}

We utilized two publicly available whole slide image datasets. The TCGA-Yale~\cite{b15} HER2 cohort consists of 182 cases of HER2 positive and negative invasive breast carcinomas. This set includes Hematoxylin and Eosin stained TCGA breast cancer slides, with slide-level annotations prepared by the Yale School of Medicine~\cite{b15}. These annotations are publicly available and show an equal distribution of 91 HER2 positive and 91 HER2 negative slides. 

We also employed the TCGA-BRCA cohort, consisting of $44$ whole slide images selected from the publicly accessible TCGA-GDC~\cite{b16} portal. These slides were part of a larger dataset of $273$ slides, from which $52$ were initially identified as having an equivocal HER2 score of 2+ based on information from cbiportal.org~\cite{b17}. HistoROI~\cite{b23}, an efficient quality control algorithm for whole slide images, was then used to process all 52 slides, filtering out those with artifacts. After this quality check, $44$ slides were selected for further analysis. The FISH test status documented by cbiportal.org for each of these slides provided reliable ground truth for categorizing them as either HER2 FISH positive or HER2 FISH negative.

\subsection{Results}

\begin{table*}[h]
\centering
\renewcommand{\arraystretch}{1.5} 
\setlength{\tabcolsep}{10pt} 
\caption{Results from ResNet50 MoCo-v2 self-supervision feature extractor for HER2 and FISH positive vs. negative classification tasks.}
\label{tab:results_table}
\begin{tabular}{|c|c|c|c|c|} 
\hline
\textbf{Task} & \textbf{Train Dataset} & \textbf{Test Dataset} & \textbf{Mean AUC ± STD} & \textbf{Max AUC} \\ \hline
HER2 Positive vs HER2 Negative  & TCGA-Yale  & TCGA-Yale & 0.85 ± 0.02  & 0.87  \\ \hline
HER2 FISH Positive vs HER2 FISH Negative  & TCGA-Yale & TCGA-BRCA & 0.81  & 0.81  \\ \hline
\end{tabular}
\end{table*}

\begin{figure}[H]
\centering
\includegraphics[width=0.5\textwidth]{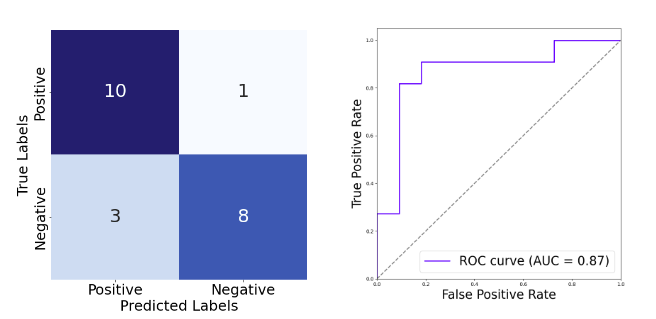}
\caption{Confusion matrix and ROC curve for HER2 positive vs HER2 negative}
\label{fig:her2posnegcmroc}
\end{figure}

This study is organized around two primary objectives, with the first objective concentrating on the classification of HER2 Positive versus HER2 Negative cases. For this task, we exclusively utilized the TCGA-Yale HER2 cohort dataset. A total of 120 whole slide images (WSIs) were carefully selected and subjected to a self-supervised learning approach using the MoCo-v2 framework. The core of this process involved training a ResNet50 encoder, which was rigorously trained over the course of 300 epochs to ensure a robust and reliable feature extraction process. Following this initial stage, the extracted features were used to train an attention module, a critical component of our classification pipeline, which was trained over an additional 100 epochs. To ensure the generalizability and reliability of our model, the dataset was systematically divided into four distinct folds. Each fold consisted of 160 slides for training purposes and 22 slides reserved for testing, allowing for a thorough evaluation of the model’s performance across different subsets of the data. During the training process, we experimented with a range of initial learning rates—specifically 1e-3, 1e-4, and 1e-5—alongside different weight decay values, also tested at 1e-3 and 1e-5. After carefully evaluating the results from all these permutations, we identified the optimal hyperparameter settings: an initial learning rate of 1e-3 coupled with a weight decay of 1e-5. These settings provided the best balance between model complexity and performance. The performance of the model was measured using the Area Under the Curve (AUC) metric, and the mean AUC achieved across the four folds was calculated to be 0.85 ± 0.02, indicating strong and consistent performance. To further illustrate the effectiveness of our approach, the confusion matrix and the Receiver Operating Characteristic (ROC) curve for the first fold are presented in Figure~\ref{fig:her2posnegcmroc}. These plots show the model’s ability to accurately distinguish between HER2 Positive and HER2 Negative cases, reflecting the overall success of our classification approach.

\begin{figure}[H]
\centering
\includegraphics[width=0.5\textwidth]{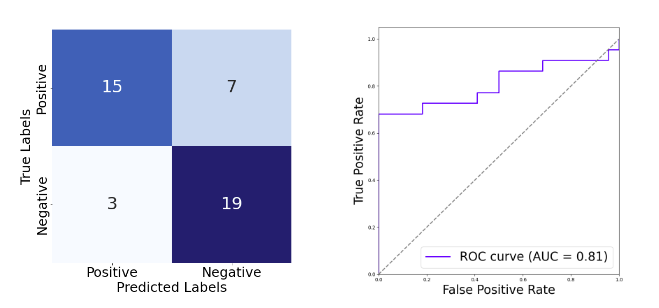}
\caption{Confusion matrix and ROC curve for HER2 FISH positive vs HER2 FISH negative}
\label{fig:fishposnegcmroc}
\end{figure}

The second major task in this study focused on classifying slides with a HER2 score of 2+, a particularly challenging category due to the ambiguity in traditional testing methods. This task involved differentiating between HER2 FISH positive and HER2 FISH negative cases, where HER2 status determination is only possible through fluorescence in situ hybridization (FISH) testing, a critical step for accurate diagnosis and treatment planning. We utilized data from both the TCGA-Yale HER2 cohort and the TCGA-BRCA cohort for this task. The ResNet50 encoder, previously trained on 120 slides in the first task, was reused to ensure consistency and leverage the learned features. For the attention module, training was conducted with 160 slides from the TCGA-Yale HER2 cohort over 100 epochs, while 44 slides from the TCGA-BRCA cohort were reserved for testing. The optimal hyperparameters from the first task—an initial learning rate of 1e-3 and a weight decay of 1e-5—were maintained. During inference on the TCGA-BRCA cohort, the model achieved an Area Under the Curve (AUC) of 0.81, demonstrating strong performance in distinguishing between HER2 FISH positive and HER2 FISH negative cases within the challenging 2+ score category. The confusion matrix and ROC curve for this task, shown in Figure~\ref{fig:fishposnegcmroc}, provide valuable insights into the model’s decision-making accuracy. 

\begin{figure*}[h]
\centering
\includegraphics[width=\textwidth]{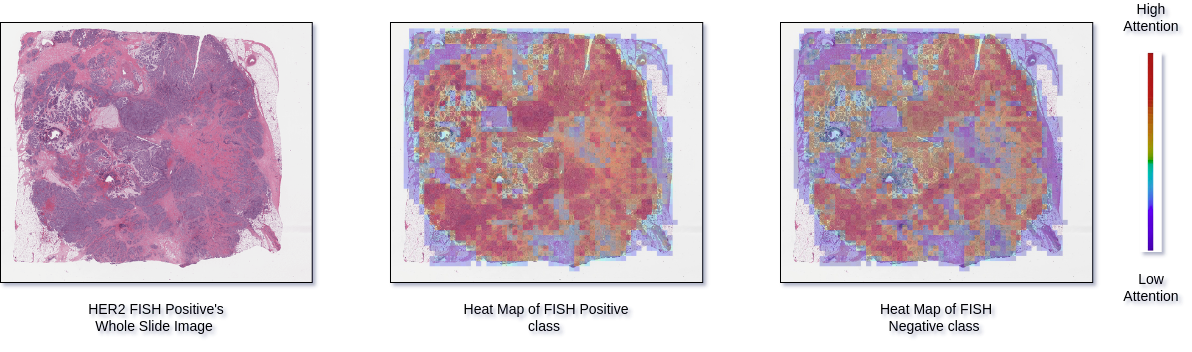}
\caption{A HER2 FISH positive whole slide image and its heatmaps}
\label{fig:positiveheatmap}
\end{figure*}

\begin{figure*}[h]
\centering
\includegraphics[width=\textwidth]{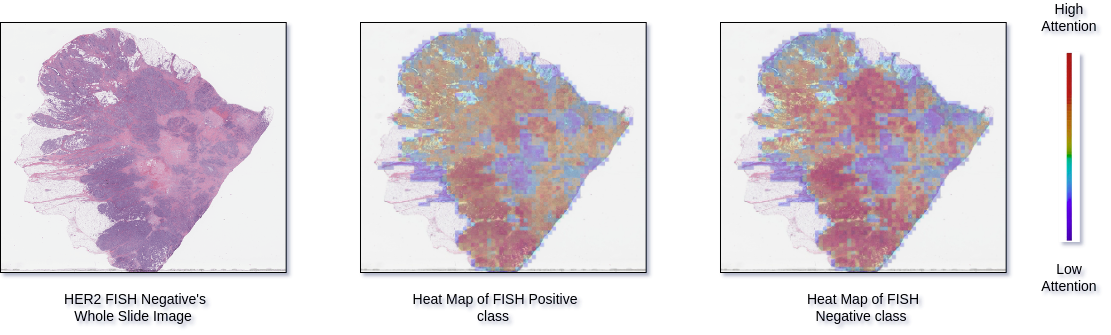}
\caption{A HER2 FISH negative whole slide image and its heatmaps}
\label{fig:negativeheatmap}
\end{figure*}
A summary of the results is presented in Table~\ref{tab:results_table}, offering a clear overview of the model's performance across both tasks. Our study's results demonstrate the effectiveness of the pipeline in predicting HER2 status. For HER2 Positive vs. Negative classification, a PPV of $0.84$ and an NPV of $0.77$ indicate robust performance in predicting both positive and negative cases. In FISH Positive vs. FISH Negative classification, a PPV of $0.83$ and an NPV of $0.73$ show strong accuracy in identifying positives, with moderate reliability for negatives. These findings highlight the potential to reduce reliance on costly IHC and FISH tests, especially in underserved populations. 



Figures~\ref{fig:positiveheatmap} and~\ref{fig:negativeheatmap} provide a detailed visualization of two whole slide images from the TCGA-BRCA dataset, each from HER2 FISH positive and FISH negative classes. For each of these slides, we generated two distinct heatmaps, corresponding respectively to the FISH positive and FISH negative classes. These heatmaps serve as visual tools to highlight the regions of the slide that the model identified as being most relevant for classification. In Figure~\ref{fig:positiveheatmap}, we focus on the HER2 FISH positive slide. The heatmaps generated for this slide reveal that the tumor regions are consistently given greater attention compared to the non-tumor areas, which underscores the model's ability to focus on clinically significant regions. Particularly, the heatmap for the FISH Positive class shows a more intense red coloration in the tumor areas compared to the Fish Negative heatmap. This vivid red coloring in the FISH Positive heatmap strongly indicates that the model has made an accurate prediction regarding the HER2 status of the entire slide. 
Conversely, Figure~\ref{fig:negativeheatmap} presents a slide that has been classified as HER2 FISH negative. In this case too, the generated heatmaps continue to emphasize the tumor regions over the non-tumor areas, indicating the model’s consistent focus on the most relevant regions for making its classification. Interestingly, the heatmap corresponding to the FISH Negative class displays a more intense red coloration in the tumor areas compared to the FISH Positive heatmap. This pattern further confirms that the model correctly predicted the HER2 status of this particular slide, demonstrating its robustness and reliability in handling complex tissue images. Overall, the heatmaps presented in these figures provide a powerful visualization of the model's decision-making process, offering insights into how the model distinguishes between HER2 FISH positive and FISH negative cases. These visualizations not only validate the model’s performance but also enhance our understanding of the underlying tissue characteristics that drive the classification outcomes.

\section{Conclusion}

In this study, we presented a novel approach for predicting HER2 status in breast cancer using hematoxylin and eosin (H\&E) stained whole slide images (WSIs). By integrating a customized weak supervision classification technique with MoCo-V2 contrastive learning, we successfully differentiated between HER2 positive and HER2 negative tumors. Our method demonstrated strong performance in both standard classification tasks and in the more challenging cases of equivocal HER2 2+ scores, which often require expensive fluorescence in situ hybridization (FISH) tests for confirmation. The application of an attention-based multiple instance learning framework allowed us to focus on the most relevant tumor regions within the WSIs, providing a more interpretable and transparent analysis of HER2 status. This approach highlights the potential of deep learning to enhance the accuracy and efficiency of HER2 status prediction using widely available H\&E stains. Our findings suggest that this computational pipeline could reduce the reliance on costly and time-consuming FISH tests, ultimately speeding up the diagnostic process and improving treatment decisions for breast cancer patients. Future research could explore incorporating additional molecular features and expanding the dataset to further enhance the model’s generalization and applicability. This study underscores the significant role of advanced computational methods in advancing personalized medicine and optimizing cancer care.

\vspace{12pt}
\end{document}